\documentclass[twocolumn,english,prl,showpacs]{revtex4}

\usepackage{graphicx}
\usepackage{esint}

\makeatletter
\@ifundefined{textcolor}{}
{%
 \definecolor{BLACK}{gray}{0}
 \definecolor{WHITE}{gray}{1}
 \definecolor{RED}{rgb}{1,0,0}
 \definecolor{GREEN}{rgb}{0,1,0}
 \definecolor{BLUE}{rgb}{0,0,1}
 \definecolor{CYAN}{cmyk}{1,0,0,0}
 \definecolor{MAGENTA}{cmyk}{0,1,0,0}
 \definecolor{YELLOW}{cmyk}{0,0,1,0}
 }

\makeatother

\usepackage{babel}

\begin{document}

\author{Izhar Neder}

\affiliation{Department of Physics, Harvard University, Cambridge, Massachusetts 02138, USA}

\title{Fractionalization noise in edge channels of integer quantum Hall states}
\begin{abstract}

A theoretical calculation is presented of current noise which is due charge
fractionalization, in two interacting edge channels in the integer
quantum Hall state at filling factor $\nu=2$. Because of the capacitive coupling between the channels, a tunneling event, in which an electron is transferred from a biased source lead to one of the two channels, generates propagating plasma mode excitations which carry fractional charges on the other edge channel. When these excitations impinge on a quantum point contact, they induce low-frequency current fluctuations with no net average current. A perturbative treatment in the weak tunneling regime yields analytical integral expressions for the noise as a function of the bias on the source. Asymptotic expressions of the noise in the limits of high and low bias are found. 
\end{abstract}
\maketitle
The fractionalization of the unit electron charge is an emergent phenomenon
which occurs in a variety of low dimensional interacting electron
systems\cite{Su1979,Laughlin1983,Pham2000,Steinberg2008}. The most
known of these is the fractional quantum Hall effect\cite{Laughlin1983},
in which the low-energy edge excitations carry a fraction of the unit
charge that can be measured by shot-noise measurements\cite{Kane1994,Saminadayar1997,dePicciotto1997}.
A different kind of such fractionalization, which is the focus of
this Letter, may also occur in the integer quantum
Hall effect (IQHE) regime. Integer quantum Hall states that have filling factors (FFs) larger
than $1$ support several copropagating edge channels\cite{Halperin1982}.
If the electrons which flow in these edge channels are strongly coupled
by Coulomb interaction, the edge excitations no longer have the usual
Fermi-liquid-like behavior. Rather, the edge channels are described
by the chiral Luttinger liquid theory, which predicts that a single
electron excitation in one edge channel separates into several copropagating
plasma modes with different velocities. Each of these modes carry
fractions of the unit electron charge  in each of the channels, depending on the interaction
strength.

These fractional charge excitations in the IQHE have not
been directly observed yet. However they may have had a crucial influence
on recent experimental results;  recently
observed energy equilibration and energy loss in the electron transport
at FF $\nu=2$\cite{Altimiras2010} suggests an energy transfer between the two channels without tunneling.  Controlled dephasing experiments of electronic interferometers\cite{Rohrlich2007,Neder2007,Roulleau2008} revealed a strong interchannel interaction at FF $\nu=2$. In addition, charge fractionalization at FF $\nu=2$ was raised as one of the explanations\cite{levkivski2008}
to the observed nontrivial behavior of the visibility of the Mach-Zehnder
interferometer (MZI) as a function of the source bias.\cite{Neder2006}

How can one measure these fractional excitations directly? The basic
idea would be to inject an electron to one channel through a tunnel barrier, and observe
the fractional excitations on the adjacent channel. Note, however,  that the fractional excitations affect neither the average current at the adjacent channel nor the low-frequency current fluctuations  - both are zero. One may try to detect the fractional charges using high frequency measurement as was proposed by Berg et. al. \cite{Berg2009,Horsdal2011}, which may be within reach with current technology \cite{Mahe2010}. 

\begin{figure}
\includegraphics[clip,scale=0.47]{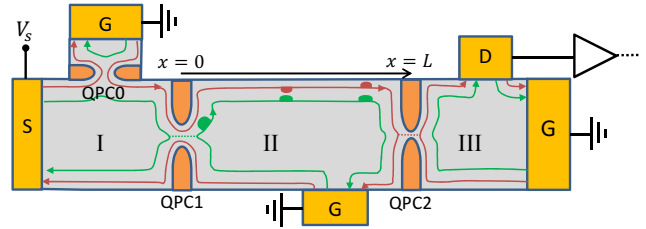} 

\caption{(color on-line) Measurement scheme.  S,D and G denote source drain and ground, respectively. The green and the red lines represent the inner and outer edge channels
in FF $\nu=2$, respectively. A tunneling of an electron in the inner channel
through QPC1 evolves into propagating slow and fast modes
with fractional positive and negative charges on both channels, as
represented by the small bulges above and below the channels lines
\label{fig:Measurement-scheme.}}

\end{figure}
In this Letter a different approach is taken, by considering a mesoscopic  device in which the fractional plasma modes impinge on a quantum point contact (QPC) which is placed on their way. The system is sketched in Fig.~\ref{fig:Measurement-scheme.}. The 2DEG bulk (the gray area) is assumed to be in a quantum Hall state with FF $\nu=2$, with two copropagating edge channels on each edge of the 2DEG. We assume that no tunneling events occur between the two channels. The source contact marked ``S'' at the top left is biased by voltage $V_{s}$ relative to other contacts, and so
it injects a net electron current $I_{s}=\frac{e^{2}}{h}V_{s}$ to each
of the two outgoing edge channels, according to Landauer formula \cite{Landauer1957,Buttiker1988}.
Those electrons propagate to the right, and impinge on QPC0, which
is tuned to selectively transmit fully only the outer channel
to a grounded contact, and completely reflect the biased inner channel
toward QPC1. QPC1 is tuned to allow small tunneling probability $T_{1}$
of electrons in the inner channel from region $\rm{I}$ to region $\rm{II}$.
An extra electron which tunnels to region $\rm{II}$ in the inner edge channel 
thus propagates along the top edge toward QPC2, presumably in a form of fractional plasma modes. QPC2, in turn, is tuned to allow
small electron tunneling probability $T_{2}$ in the \emph{outer
channel }from region $\rm{II}$ to region $\rm{III}$. A tunneling event through
QPC2 creates an extra charge on the outer edge channel in region $\rm{III}$,
which propagates to the contact at the top right of Fig.~1 and adds temporarily to the current which is picked up by an amplifier.

In the above setup, the fractionalization
of the extra electrons in region $\rm{II}$ leads
to low-frequency current noise in region $\rm{III}$. Below, I first present a qualitative argument for the existence of this noise. The noise is then calculated as a function the source bias  $V_s$ at zero temperature in the weak tunneling limit of the QPCs, using the chiral Luttinger liquid theory. It is found that the noise behavior has a crossover; in the limit of large source bias the noise scales as $V_s^{1+2\eta}$ where $0<\eta<0.5$ depends on the coupling between the channels (see definition below). In the low bias limit the noise vanishes as fast as $V_s^3$.

The appearance of the noise can be explained intuitively as follows. Suppose that an electron tunnels to the inner channel in region $\rm{II}$ through QPC1. Because of the interaction between the two channels, the extra charge breaks up into two density modes,
both propagating to the right; a fast, ``charge-like''
mode, with negative fractional charge excitations on both edge
channels, and a slow, ``dipole-like'' mode, with extra
negative charge on the inner channel and extra positive charge (holes)
on the outer channel (see Fig.~\ref{fig:Measurement-scheme.}). When
each mode arrives to be near QPC2, it allows temporarily a certain
tunneling event from region $\rm{II}$ to region $\rm{III}$, in the outer edge channel.
The charge-like mode induces a tunneling probability of an electron
above the Fermi level, while the dipole-like mode induces a tunneling
probability of a hole below the Fermi level. The two density modes have different velocities, and so
they arrive to QPC2 at different times, which are well resolved if the uncertainty in the energy  of the tunneling electron is large enough (i.e.~for high enough source bias). In this case the two corresponding tunneling events are  separated in time and are statistically uncorrelated. As a result, the net extra
current in region $\rm{III}$ is expected to be zero on average, with an equal
average number of electrons and holes tunneling through QPC2. However
as the tunneling events of electrons and holes are random, the current
will fluctuate around the zero average, and these fluctuations will
have a low-frequency component, similar to the usual Schottky noise
from a tunnel barrier.

For a quantitative prediction for the noise,  let us model the system by a low-energy effective theory in the lowest Landau level, using the Hamiltonian 
\[
H=\sum_{R}H_{R}+H_{QPC1}+H_{QPC2}.\]
 The index $R$ goes over the regions, $R\in\{\rm{I,II,III}\}$.
$H_R$ describes the evolution in the two edge channels within region R, corresponding to electrons with spin up and spin down relative to the magnetic field direction, with Coulomb interaction between the channels. In an appropriate choice of gauge, one has ($\hbar=1$)
\begin{eqnarray}
H_{R} & = & -i\int_{-\infty}^{\infty} dx\left(v_{\rm{in}}\psi_{\rm{in},R}^{\dagger}\frac{\partial}{\partial x}\psi_{\rm{in},R}+v_{\rm{out}}\psi_{\rm{out},R}^{\dagger}\frac{\partial}{\partial x}\psi_{\rm{out},R}\right)\nonumber \\
 &  & +\frac{u}{2\pi}\int_{-\infty}^{\infty} dx:\rho_{\rm{in},R}\rho_{\rm{out},R}:+\delta_{R,1}eV_s\int_{-\infty}^{\infty} dx:\rho_{\rm{in},R}:. \label{eq:H_R}\end{eqnarray}
Here $\psi_{\rm{in}(out),R}\left(x,t\right)$ is the electron annihilation
operator in region $R$ in the inner (outer) channel (the Heisenberg picture is used throughout the Letter) and $\rho_{\rm{in}(out).R}\left(x,t\right)\equiv\psi_{\rm{in}(out),R}^{\dagger}\psi_{\rm{in}(out),R}$
are the 1d electron number densities at the channels. ':' denotes normal ordering, $v_{\rm{in}}$ and
$v_{\rm{out}}$ are the bare velocities of the two channels and $u$ is
the interaction strength. The grounded contacts are modeled effectively by setting the integral boundaries to $\pm\infty $. The last term in Eq.~(\ref{eq:H_R}) models the bias $eV_S$ of the inner channel in region $\rm{I}$ after QPC0.

$H_{QPC1}$ describes the tunneling at QPC1, at $x=0$, between the
inner channels in regions $\rm{I}$ and $\rm{II}$. $H_{QPC2}$ describes the
tunneling at QPC2, at $x=L,$ between the outer channels of regions
$\rm{II}$ and $\rm{III}$. They are given by \begin{equation}
H_{QPC1}=\bar{v}_{\rm{in}}\sqrt{T_{1}}\psi_{\rm{in},II}^{\dagger}(0)\psi_{\rm{in},I}(0)+\rm{H.c}.\label{eq:H_QPC1}\end{equation}
 \begin{equation}
H_{QPC2}=\bar{v}_{\rm{out}}\sqrt{T_{2}}\psi_{\rm{out},III}^{\dagger}(L)\psi_{\rm{out},II}(L)+\rm{H.c.}\label{eq:H_QPC2}\end{equation}
 Here $T_{1}$ and $T_{2}$ are the electron transmission probabilities
of QPC1 and QPC2, respectively. $\bar{v}_{\rm{in}}$ and $\bar{v}_{\rm{out}}$
are the renormalized tunneling density of states in the inner and outer channels,
which are  assumed here for simplicity to be equal for all three regions.
They cancel out in the calculation below and do not appear
in the final formula for the noise.

The measured noise in region $\rm{III}$ is given by \cite{Martin2005}
\begin{equation}
S_{f\rightarrow0}=\int_{-\infty}^{\infty}dt\left\langle \Phi_{eV_{s}}\right|\left\{ I_{\rm{out},III}(t),I_{\rm{out},III}(0)\right\} \left|\Phi_{eV_{s}}\right\rangle. \label{eq:noise1}\end{equation}
 The state $\left|\Phi_{eV_{s}}\right\rangle $ is the ground state
of the unperturbed Hamiltonian $H_{R}$ (i.e. with no tunneling events
between the various regions), where the inner channel of region $\rm{I}$
is biased by $eV_{s}$ and all other edge channels are grounded. The
current operators in Eq.~(\ref{eq:noise1}), $I_{\rm{out},III}\equiv e\frac {d}{dt}\int_{-\infty}^{\infty} dx \rho_{\rm{out},III}$, can be written using the
tunneling operators,
\begin{equation}
I_{\rm{out},III}(t)=i\bar{v}_{\rm{out}}\sqrt{T_{2}}\left[\psi_{\rm{out},III}^{\dagger}(L,t)\psi_{\rm{out},II}(L,t)-\rm{H.c.}\right].\label{eq:I_t}\end{equation}
The noise is calculated by expanding the time evolution of the current operators
to first order in each of the tunneling probabilities $T_{1}$ and $T_{2}$
using Keldysh formalism\cite{Keldysh1964,Rammer1986}. 

The evolution in region $\rm{II}$ is solved analytically. Note that given the Hamiltonian
$H_{R=II}$ in Eq.~(\ref{eq:H_R}), The equation of motion for the density is
given by
\begin{equation}
\frac{\partial}{\partial t}\left(\begin{array}{c}
\rho_{\rm{in},II} \\ \rho_{\rm{out},II}
\end{array}\right)
+U\frac{\partial}{\partial x}\left(\begin{array}{c}
\rho_{\rm{in},II} \\ \rho_{\rm{out},II}
\end{array}\right)=0,
\label{eq:eom}\end{equation}
 where U= $\left(\begin{array}{cc}
v_{\rm{in}} & u\\
u & v_{\rm{out}}\end{array}\right)$ is the velocity matrix. The density modes of the dynamics
are the eigenstates of $U$, which can be written in an orthonormal basis  as $\left(\begin{array}{c}
\cos\theta\\
\sin\theta\end{array}\right)$ and $\left(\begin{array}{c}
\sin\theta\\
-\cos\theta\end{array}\right)$, where $0<\theta<\pi/2$. The velocities of the density modes
are the eigenvalues of U,
\begin{equation}
v_{1,2}=\frac{v_{\rm{in}+}v_{\rm{out}}}{2}\pm\sqrt{\frac{1}{4}\left(v_{\rm{in}}-v_{\rm{out}}\right)^{2}+u^{2}},\label{eq:v_12}\end{equation}
where the $+(-)$ sign refers to the fast(slow) mode.

The noise in Eq.~(\ref{eq:noise1}) is now calculated in two steps. First, we expand the time evolution of the current operators to second order in $\sqrt{T_{2}}$. Using Fourier transform to express the result as integral over
energies, one finds
\begin{equation}
S_{f\rightarrow0}=2g_0T_{2}\int_{0}^{\infty}d\omega n_{L}(\omega)+2g_0T_{2}\int_{-\infty}^{0}d\omega\left(1-n_{L}(\omega)\right),\label{eq:noise11}\end{equation}
where $g_0=\frac{e^{2}}{h}$ is the unit conductance, and the function $n_{L}(\omega)$ is given by the correlator
\begin{equation}
n_{L}(\omega)=\bar{v}_{\rm{out}}\left\langle \Phi_{eV_s}\right|\psi_{\rm{out},II}^{\dagger}(L,0)\psi_{\rm{out},II}(L,t)\left|\Phi_{eV_s}\right\rangle _{\rm{FT}},\label{eq:n_L_w}\end{equation}
 where ``FT'' denotes Fourier transform. $n_{L}(\omega)$ can be interpreted as the effective mean occupation
of electron states at energy $\omega$ in the outer channel in region $\rm{II}$
at $x=L$, near QPC2. Without inter-channel interaction, in the
case $u=0$, $n_{L}(\omega)$
would be a Fermi distribution at zero temperature, which is a step
function $n_{L}(\omega)=\Theta\left(-\omega\right)$. However, when $u\neq0$,
the noisy inner channel excites the electrons in the outer channel and changes
its occupation distribution function. One can write
\begin{equation}
n_{L}(\omega)=\int_{-\infty}^{\infty} \frac{d\omega_{\rm{out}}}{2\pi}\Theta\left(-\omega+\omega_{\rm{out}}\right)B\left(\omega_{\rm{out}}\right). \label{eq:n_w}\end{equation}
The function $B\left(\omega_{\rm{out}}\right)$ is related to the probability for an
electron-hole excitation with energy $\omega_{\rm{out}}$ in the outer
channel in region $\rm{II}$. In the absence of net current in the outer channel, the
electrons and the holes contribute equally to the noise. We can therefore consider only the first term
in Eq.~(\ref{eq:noise11}) twice, which leads to
\begin{equation}
S_{f\rightarrow0}=4g_0T_{2}\int_{0}^{\infty}\frac{d\omega_{\rm{out}}}{2\pi}\omega_{\rm{out}}B(\omega_{\rm{out}}).\label{eq:noise2}\end{equation}
The second step is to calculate the function $B(\omega_{\rm{out}})$ up to second order in the transmission
amplitude of QPC1, $\sqrt{T_{1}}$, using the chiral Luttinger liquid theory (see supplementary material for details).  One finds
\begin{equation}
B(\omega_{\rm{out}})=T_{1}\tau\int_{0}^{e\tilde{V_{s}}}\frac{d\tilde{\omega}_{\rm{in}}}{2\pi}\left(e\tilde{V_{s}}-\tilde{\omega}_{\rm{in}}\right)C\left(\tilde{\omega}_{\rm{out}},\tilde{\omega}_{\rm{in}}\right).\label{eq: Bw}\end{equation}

Here $\tau=\frac{L}{v_{2}}-\frac{L}{v_{1}}$ is the relative delay
of the arrival of the two modes from QPC1 to QPC2. We define the second
mode to be the dipole-like mode, such that $\tau>0$. In Eq.~(\ref{eq: Bw})
we also have $e\tilde{V_{s}}=eV_{s}\tau$, and $\tilde{\omega}_{\rm{out}}=\omega_{\rm{out}}\tau$.

\begin{figure}
\includegraphics[scale=0.35]{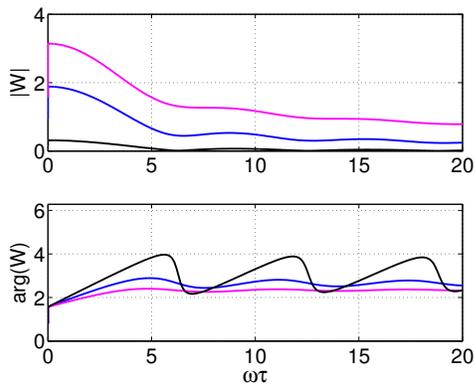} \caption{ (color online) $W(\tilde{\omega})$, absolute value and phase, for $\eta=0.5$ (magenta),
$\eta=0.3$ (blue) and $\eta=0.05$ (black)\label{fig: W_w} }
\end{figure}

The function $C\left(\tilde{\omega}_{\rm{out}},\tilde{\omega}_{\rm{in}}\right)$ weights the contribution of processes with energy loss $\tilde{\omega}_{\rm{in}}/\tau$ in the inner channel to the probability of electron-hole excitations with energy $\tilde{\omega}_{\rm{out}}/\tau$ in the outer channel. Note, however, that this function can have negative values. It does
not depend on the bias, and is a property of the free evolution of
the two modes from $x=0$ to $x=L$. 
It is found to be a sum
of three terms, $C\left(\tilde{\omega}_{\rm{out}},\tilde{\omega}_{\rm{in}}\right)=C^{(2)}+C^{(3)}+C^{(4)}$, with 
\begin{eqnarray}
C^{(2)}&=&\left(2\pi\right)\left|W\left(\tilde{\omega}_{\rm{out}}\right)\right|^{2}\delta\left(\tilde{\omega}_{\rm{in}}-\tilde{\omega}_{\rm{out}}\right)\label{eq:c2}\\
C^{(3)}&=&2\Re\left\{e^{i(-\tilde{\omega}_{\rm{in}}+\tilde{\omega}_{\rm{out}})}W\left(\tilde{\omega}_{\rm{in}}\right)W\left(\tilde{\omega}_{\rm{in}}-\tilde{\omega}_{\rm{out}}\right)W^{*}\left(\tilde{\omega}_{\rm{out}}\right)\right\} \nonumber\\
C^{(4)}&=&e^{i(\tilde{\omega}_{\rm{in}}+\tilde{\omega}_{\rm{out}})}\int\frac{d\omega}{2\pi}e^{-i2\tilde{\omega}}W\left(\tilde{\omega}\right)W^{*}\left(\tilde{\omega}_{\rm{in}}-\tilde{\omega}\right)\nonumber\\
&&\times W\left(-\tilde{\omega}_{\rm{out}}+\tilde{\omega}\right)W^{*}\left(\tilde{\omega}_{\rm{in}}-\tilde{\omega}+\tilde{\omega}_{\rm{out}}\right),\label{eq:c4}
\end{eqnarray}
where $W\left(\tilde{\omega}\right)=w(\tilde{t}){}_{FT}$,
and
\begin{equation}
w(\tilde{t})=\frac{(\tilde{t}+1+i\delta)^{\eta}}{(\tilde{t}+i\delta)^{\eta}}-1.\label{eq:w_t}\end{equation}
The power $\eta$ is related to the density modes of Eq.~(\ref{eq:eom}), $\eta=\cos{\theta}\sin{\theta}=4u/\sqrt{\left(v_{1}-v_{2}\right)^{2}+4u^{2}}$. Thus, the function $W\left(\tilde{\omega}\right)$ is directly related to the fractionalization effect. 
Fig.~(\ref{fig: W_w}) shows the function $W\left(\tilde{\omega}\right)$
for three possible values of the power $\eta$. Note that it vanishes at negative values, and satisfies $W(0^+)=2\pi i\eta$.
Also note that the power-law
tail of $W\left(\tilde{\omega}\right)$ at high energies is directly related
to the power-law divergence of $w(\tilde{t})$ at $t=0$. The asymptotic behavior is  
$W\left(\tilde{\omega}\gg1\right)\approx\frac{2\pi\eta}{\Gamma(1+\eta)}\omega^{\eta-1}$, where $\Gamma$ is the gamma function.

\begin{figure}
\includegraphics[scale=0.35]{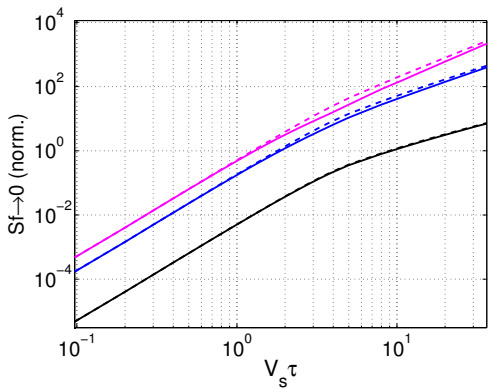}
\caption{ (color on-line) The fractionalization noise, $S_{f\rightarrow0}/(2g_0T_{1}T_{2}\tau^{-1})$
as a function of $\tilde{eV_{s}}=eV_{s}\tau$, for $\eta=0.5,0.3,0.05$ (magenta, blue and black solid lines, respectively).
The corresponding dashed lines show the contribution of the term $C^{(2)}$ to the noise for the same values of $\eta$. \label{fig:the-fractionalization-noise}}
\end{figure}

The fractionalization noise $S_{f\rightarrow0}$, calculated according to Eqs.~(11)-(15), is plotted in Fig.~\ref{fig:the-fractionalization-noise} as a function of $e\tilde{V}_{s}$.
The contribution of the noise only from $C^{(2)}$ is also plotted in Fig.~\ref{fig:the-fractionalization-noise}. %
In the limit of small bias, $e\tilde{V}_{s}\ll1$, the leading order of the noise comes from the contribution of $C^{(2)}$ at small values of  $\tilde{\omega}_{\rm{in}}$, that is from the value of $|W\left(0^+\right)|^2$.  Equations (\ref{eq:noise2}) and (\ref{eq: Bw})
then lead to 
\begin{equation}
S_{f\rightarrow0}|_{\tilde{V}_{s}\ll1}\approx\frac{4\pi\eta^{2}}{3}g_{0}T_{1}T_{2}\tau^{-1}\left(e\tilde{V_{s}}\right)^{3}.
\end{equation}
In the opposite limit,
of $e\tilde{V}_{s}\gg1$, The noise is a power-law function  of the bias. One can see from Fig.~\ref{fig:the-fractionalization-noise} that
still the main contribution to the noise comes from $C^{(2)}$. In this limit
the noise is dominated by the tail of $\left|W\left(\tilde{\omega}_{\rm{out}}\right)\right|^{2}$
at high $\tilde{\omega}_{\rm{in}}$. Approximating $C\approx C^{(2)}$ and using again Eq.~(\ref{eq:c2})
in Eqs.~(\ref{eq:noise2}) and (\ref{eq: Bw}),  one finds for this limit 
\begin{equation}
S_{f\rightarrow0}\approx2g_0T_{1}T_{2}\tau^{-1}\frac{2\pi\eta}{\left(2\eta+1\right)\left[\Gamma(1+\eta)\right]^{2}}\left(e\tilde{V}_{s}\right)^{1+2\eta}.
\end{equation}
In summary, the low-frequency noise due to the fractionalization
effect was calculated in the integer quantum Hall effect at FF $\nu=2$ at zero temperature.
The noise is a result of the fractional density modes impinging on
a QPC and inducing excess tunneling events of electrons and holes through the
QPC to the outgoing leads. The fractionalization noise is found to vanish faster at the low source bias limit, where the time of arrival of the two density modes to QPC2 is no longer well resolved. Thus, the bias in which the crossover occurs in the behavior of the noise corresponds to the difference in the arrival times of the modes from QPC1 to QPC2, $eV_s|_{\rm{crossover}}\approx\hbar\tau^{-1}$. This crossover bias may be roughly estimated, based on energy scales which appeared in recent experimental results \cite{Altimiras2010,Neder2007,Roulleau2008}. If they are indeed related to the same fractionalization effect, then the energy scale is at the order $\approx10\,\mu eV$ for a device with typical length of $10\,\mu m$. The crossover voltage  should be different for devices with different typical lengths and different edge profiles.
Finally, it should be mentioned that the effect of finite temperature and the effect of the disorder in the edge channels in region
II on the behavior of the noise were not discussed in the model above and deserves future study.

I acknowledge B. I. Halperin, G. Viola, Y. Oreg and E. Berg for very useful discussions. This work was supported by NSF grant DMR-0906475. 
\bibliography{frac_bib}
 
\end{document}


\author{Izhar Neder}
\affiliation{Department of Physics, Harvard University, Cambridge, MA 02138, USA}

\title{Fractionalization noise in edge channels of integer quantum Hall states - on-line supporting material}

\maketitle

\subsection*{Details of the perturbative calculation of the noise using the chiral Luttinger  liquids theory }

The derivation from Eq.~(4) to Eq.~(8) in the main text is via perturbation expansion in the parameter $\sqrt{T_2}$ at zero temperature, and is rather straightforward - see for example Ref. 16 for similar derivations. The derivation from Eq.~(8) to Eqs.~(12)-(15) in the main text is less obvious and involves some aspects of the chiral Luttinger liquid theory, as is elaborated here.  

Using Keldysh technique to expand $n_{L}\left(\omega\right)$ in Eq.~(9) in the main text to second order in the transmission amplitude $\sqrt{T_{1}}$, one finds 
\begin{widetext}
\begin{equation}
n_{L}(w)=\Theta\left(-\omega\right)+\bar{v}_{in}\bar{v}_{out}T_{1}\left[\int_{-\infty}^{eV_{s}}\frac{d\omega_{in}}{2\pi}G^{(e)}(\omega,\omega_{in})+\int_{eV_{s}}^{\infty}\frac{d\omega_{in}}{2\pi}G^{(h)}(\omega,\omega_{in})\right],\label{eq:nL2}\end{equation}

where 
\begin{equation}
G^{[e(h)]}(\omega,\omega_{in})=\iiintop_{-\infty}^{\infty}dtdt_{1}dt_{2}g^{[e(h)]}(t,t_{1},t_{2})e^{-iwt-iw_{in}\left(t_{2}-t_{1}\right)}
\label{eq:G_g}\end{equation}
and the two correlators $g^{[e(h)]}(t,t_{1},t_{2})$
are given by
\begin{eqnarray}
g^{(e)}(t,t_{1},t_{2}) & = & \left\langle \Phi_{eV_s}\right|\psi^0_{in,\rm{II}}(0,t_{2})\psi_{out,\rm{II}}^{0\dagger}(L,t)\psi^0_{out,\rm{II}}(L,0)\psi_{in,\rm{II}}^{0\dagger}(0,t_{1})\left| \Phi_{eV_s}\right\rangle \nonumber, \\
g^{(h)}(t,t_{1},t_{2}) & = & \left\langle\Phi_{eV_s}\right|\psi_{in,\rm{II}}^{0\dagger}(0,t_{1})\psi^{0\dagger}_{out,\rm{II}}(L,t)\psi^0_{out,\rm{II}}(L,0)\psi^0_{in,\rm{II}}(0,t_{2})\left|\Phi_{eV_s}\right\rangle .\label{eq:g_e}\end{eqnarray}
\end{widetext}

 The superscript ``0'' on the $\psi$ operators denotes an evolution according to the free Hamiltonian $H_{\rm{II}}$ in Eq.~1 in the main text. Note that all the other correlators
appearing the Keldysh perturbation calculation of $n_{L}\left(\omega\right)$,
which have different ordering of the four $\psi$ operators, vanish.
This was checked for each such correlator by performing the integration over $t$ in the FT in Eq.~(9) in the main text using contour integration in the complex t plane, and rearranging the outcome. 

Only the integral term in the r.h.s.~of Eq.~(\ref{eq:nL2}) contribute
to the noise $S_{f\rightarrow0}$.  The derivation of the integral of $G^{(e)}\left(\omega,\omega_{in}\right)$   is given here in details. The integral of  $G^{(h)}\left(\omega,\omega_{in}\right)$ can be calculated by similar steps and gives no contribution at all. The physical reason for that is that at zero temperature we injects only electrons to region II through QPC1, and not holes. 

Following Eqs.~(\ref{eq:G_g}) and (\ref{eq:g_e}), the correlator $g^{(e)}(t,t_{1},t_{2})$ is calculated
using bosonization technique, by expressing the electron fields operators with
the boson fields of the two chiral modes,

\begin{eqnarray}
\psi^0_{in,\rm{II}}(x,t) & = & \sqrt{\frac{\Lambda}{2\pi}}e^{i\left[cos\theta\phi_{1}(x-v_{1}t)+sin\theta\phi_{2}(x-v_{2}t)\right]}\nonumber \\
\psi^0_{out,\rm{II}}(x,t) & = & \sqrt{\frac{\Lambda}{2\pi}}e^{i\left[sin\theta\phi_{1}(x-v_{1}t)-cos\theta\phi_{2}(x-v_{2}t)\right]},\label{eq:psi_phi}\end{eqnarray}

where $\Lambda$ is the momentum cutoff. It is further assumed here that due to the presence of the grounded ohmic contacts, any finite size effects which result from boundary conditions in the  x axis  can be neglected. The two boson fields thus satisfy
the commutation relations

\begin{eqnarray}
[\phi_{i}(x),\phi_{j}(x')] & = & \pi i\delta_{ij}{\rm {\rm sign}(x-x')}\label{eq:Comm_rel}\end{eqnarray}

  and satisfy the diagonalized equations of motion,

\[
\dot{\phi_{i}}=v_{i}\frac{\partial\phi_{i}}{\partial x},\, i=1,2,\]
where $v_{1},v_{2}$ are the velocities of the fast and slow modes  which are
given in Eq.~(7) in the main text. 
 The expressions in Eq.~(\ref{eq:psi_phi}) is then inserted into Eq.~(\ref{eq:g_e}), which is then calculated by comparing its r.h.s to the l.h.s of  the following identity: for any four operators of a boson field $\{\phi(x_j)\},j=1,...,4$ and a set of constant
$a_{1}...,a_{4}$ with vanishing sum, $\sum_{k}a_{k}=0$, one has

\begin{equation}
\left\langle \Phi_{eV_s}\right|e^{ia_{1}\phi(x_{1})}\cdot...\cdot e^{ia_{4}\phi(x_{4})}\left|\Phi_{eV_s}\right\rangle   =  \prod_{j<k}e^{-a_{j}a_{k}K(x_{j}-x_{k})},\label{eq:a1__a4}\end{equation}

with $K(x_{j}-x_{k})=\left\langle \Phi_{eV_s}\right|\left[\phi(x_{j})-\phi(x_{k})\right]\phi(x_{k})\left|\Phi_{eV_s}\right\rangle $.
Eq. (\ref{eq:a1__a4}) is proved by repeatedly using the Baker-Hausdorff
formula $e^{iA}e^{iB}=e^{i\left(A+B\right)}e^{-\frac{1}{2}[A,B]}$
four times, and then using the identity of boson fields $\left\langle \Phi_{eV_s}\right|e^{i\sum_{k}a_{k}\phi_{i}(x_{k})}\left|\Phi_{eV_s}\right\rangle =e^{-\frac{1}{2}\left\langle \Phi_{eV_s}\right|\left[\sum_{k}a_{k}\phi_{i}(x_{k})\right]^{2}\left|\Phi_{eV_s}\right\rangle }$. 

Equations (\ref{eq:g_e}) + (\ref{eq:psi_phi}) leads to a product of  two correlators similar to the one in the l.h.s. of Eq.~(\ref{eq:a1__a4}), one for
each boson mode, $\phi=\phi_{1}$ and $\phi=\phi_{2}$,  with different parameters
$a_{1},..,a_{4}$: In the correlator for $\phi_{1}$ the parameters are $a_{1}=-\cos\theta,\ a_{2}=-\sin\theta,\ a_{3}=\sin\theta,\ a_{4}=\cos\theta$.
In the correlator for $\phi_{2}$ the parameters are $a_{1}=-\sin\theta,\ a_{2}=\cos\theta,\ a_{3}=-\cos\theta,\ a_{4}=\sin\theta$.
For both fields, given Eq.~ (\ref{eq:Comm_rel}), the correlator
$K$ satisfies $\frac{\Lambda}{2\pi}e^{K(x)}=\frac{1}{x+i\delta}$. Applying the identity in Eq.~(\ref{eq:a1__a4}) results in a product over all possible pairing of bosons operators, which can be written as

\begin{equation}
g^{(e)}(t,t_{1},t_{2})=g_{out}^{<}(-t)g_{in}^{>}(t_{2}-t_{1})c(t,t_{1},t_{2})\label{eq:g_e2}.\end{equation}

The r.h.s.~of Eq.~(\ref{eq:g_e2}) is a product of three functions. The first two are the lesser and greater correlators, which result from the all the pairings of boson operators whose origin are two $\psi$ operators of the same edge channel (either the inner or the outer channel). They are given by   

\begin{eqnarray}
g_{in}^{>}(t) & = & -i\left\langle \Phi_{eV_s}\right|\psi^0_{in,\rm{II}}(0,t)\psi_{in,\rm{II}}^{0\dagger}(0,0)\left|\Phi_{eV_s}\right\rangle =\nonumber\\
 & = & -\frac{1}{2\pi\bar{v}_{in}}\cdot\frac{1}{t+i\delta}\nonumber\\
g_{out}^{<}(t) & = & i\left\langle \Phi_{eV_s}\right|\psi_{out,\rm{II}}^{0\dagger}(L,0)\psi^0_{out,\rm{II}}(L,t)\left|\Phi_{eV_s}\right\rangle \nonumber\\
 & = & -\frac{1}{2\pi\bar{v}_{out}}\cdot\frac{1}{t-i\delta},\label{eq:g_gl}\end{eqnarray}

where $\bar{v}_{in}=v_{1}^{\cos^{2}\theta}v_{2}^{\sin^{2}\theta}$ and $\bar{v}_{out}=v_{1}^{\sin^{2}\theta}v_{2}^{\cos^{2}\theta}$. One should note here that the above correlators $g^{>}$ and $g^{<}$ are also the ones that appear in the expression for the current through a biased QPC. Therefore  $\bar{v}_{in}$ and $\bar{v}_{out}$ are identified with the
renormalized tunneling densities of states at the QPCs which appear in Eqs.~(2) and (3) in the main text. 

The function $c(t,t_{1},t_{2})$ in Eq.~(\ref{eq:g_e2}) results
from the pairing of boson operators whose origin are two $\psi$ operators  from different edge channels. There are four such pairing possibilities, hence one finds
\begin{widetext}
\begin{equation}
c(t,t_{1},t_{2})  =  \left[1+w^{*}(\frac{t_{1}+\tau_{1}}{\tau})\right]\left[1+w(\frac{t-t_{1}-\tau_{2}}{\tau})\right]\left[1+w(\frac{-t+t_{2}+\tau_{1}}{\tau})\right] \left[1+w^{*}(\frac{-t_{2}-\tau_{2}}{\tau})\right],\label{eq:wx4}\end{equation}

where the function $w\left(x\right)$ is given in
Eq.~(15) in the main text, and where $\tau_{1}=L/v_{1}$, $\tau_{2}=L/v_{2}$
and $\tau=\tau_{2}-\tau_{1}$. Note that for noninteracting case, $w=0$ and hence $c(t,t_{1},t_{2})=1$.

Equations (\ref{eq:nL2}),  (\ref{eq:G_g}) and (\ref{eq:g_e2})-(\ref{eq:wx4}) leads to Equations (10)+(12) in the main text; when  inserting equations  (\ref{eq:g_e2})-(\ref{eq:wx4}) into Eq.~(\ref{eq:G_g}), performing the Fourier transform in  Eq.~(\ref{eq:G_g}) and inserting the result into the integral term in Eq.~(\ref{eq:nL2}), one gets equations (10)+(12) in the main text, with  the function $C\left(\tilde{\omega}_{in},\tilde{\omega}_{out}\right)$ given by 
\begin{equation}
C\left(\tilde{\omega}_{in},\tilde{\omega}_{out}\right)=\frac{1}{\tau^{3}}\int_{-\infty}^{\infty}dtdt_{1}dt_{2}c(t,t_{1},t_{2})e^{-i\tilde{\omega}_{out}t/\tau-i\tilde{\omega}_{in}\left(t_{2}-t_{1}\right)/\tau}\label{eq:C_w2}.\end{equation}
\end{widetext}
Note that $C\left(\tilde{\omega}_{in},\tilde{\omega}_{out}\right)$ gets contribution from terms in the expansion of Eq.~(\ref{eq:wx4}) that contains product of two, three and four functions $w(x)$. These correspond to $C^{(2)}$, $C^{(3)}$ and $C^{(4)}$ in the main text. The contribution of terms containing only one $w(x)$ vanishes. In this context, note that the calculation is greatly simplified by the fact that  the Fourier transform of $w(x)$ vanishes for negative frequencies, which causes the contributions to the noise of most of the terms in the expansion to vanish. 